\documentstyle[aps,multicol,graphicx,amsmath,amsfonts,psfrag]{revtex}
\graphicspath{{figs/}}
\newcommand{\br}{{\mbox{$\mathbf{r}$}}}
\newcommand{\bp}{{\mbox{$\mathbf{p}$}}}

\newcommand{\be}{\begin{equation}}
\newcommand{\ee}{\end{equation}}

\newcommand{\ri}{{\rm i}}
\newcommand{\ds}{\displaystyle}
\begin{document}
\title{Signature of ballistic effects in disordered conductors}
\author{Ville Uski$^1$, Bernhard Mehlig$^2$ and
Michael Schreiber$^1$}
\address{
\mbox{}$^1$Institut f\"ur Physik, Technische Universit\"at Chemnitz,
           D-09126 Chemnitz, Germany\\
\mbox{}$^2$Complex Systems, Physics \& Engineering Physics, CHALMERS TH/GU, 41296 Gothenburg, Sweden
        }
\date{\today}
\maketitle{ } 
\begin{abstract}
Statistical properties of energy levels, wave functions and 
quantum-mechanical matrix elements in disordered conductors
are usually calculated assuming
diffusive electron dynamics.
Mirlin  has pointed out [Phys. Rep. {\bf 326},  259  (2000)]
that ballistic effects
may, under certain circumstances, dominate
diffusive contributions. 
We study the influence of such ballistic effects
on the statistical properties of wave functions
in quasi-one dimensional disordered conductors.
Our results support the view  that 
ballistic effects 
can be significant in
these systems.   
\end{abstract}
\begin{multicols}{2}

One of the simplest and yet most widely used
models \cite{LifGP88,Efe97,Mir00} for a disordered wire is that
of  independent electrons
moving in a disordered potential
described by a $\delta$-correlated 
Gaussian random function 
$\upsilon(\br)$ (where $\br=(x,y,z)$
is a spatial coordinate)
\begin{equation}
\label{eq:corr}
\begin{split}
\langle \upsilon(\mathbf{r}) \rangle &= 0\,,\\
\langle \upsilon(\mathbf{r}) \upsilon(\mathbf{r}')\rangle
&= (2\pi\nu\tau)^{-1} \delta(\mathbf{r}-\mathbf{r}')\,.
\end{split}
\end{equation}
Here $\langle \cdots\rangle$ is an average
over potential realisations, 
$\nu=1/(V\Delta)$ the electronic density of states,
$V$ the volume of the system, $\Delta$ the mean level spacing, and $\tau$ the scattering time. 
The corresponding Hamiltonian is
\begin{equation}
\label{eq:H}
 H =  \frac{\bp^2}{2m} + \upsilon(\br)
\end{equation}
with mass $m$ and momentum $\bp$.
For the model (\ref{eq:corr},\ref{eq:H}) 
it has been proven \cite{Efe97}
that on small energy scales (of 
the order of $\Delta$) and
for sufficiently weak disorder the
quantum-mechanical correlation
functions are universal and equivalent to those of random matrix theory (RMT)
\cite{meh67,por65}. 
Moreover, deviations from universal statistics
are parameterised by the
dimensionless conductance
$g = 2\pi\hbar  D /(L^2\Delta)$ where
$D$ is the diffusion constant and $L$
is the linear extension of the system.
In the limit of $g\rightarrow\infty$,
fluctuations are of RMT type.  Deviations may be calculated perturbatively,
as an expansion in $g^{-1}$ (see for instance
\cite{Efe97,Mir00} and references therein).

These important results were established
using an effective field theory, the 
nonlinear $\sigma$ model (NLSM) \cite{Efe97}, valid
for $ \lambda_{\rm F} \ll \ell$ (where
$\lambda_{\rm F}$ is the Fermi wave length) 
and on length scales larger than the mean free
path $\ell = v_{\rm F}\tau$ 
(where $v_{\rm F}$ is the Fermi velocity).
Efetov's NLSM assumes that the electrons undergo 
diffusive motion and, within a saddle-point
approximation, statistical
properties of eigenvalues, wave functions,
and matrix elements are characterised
by a classical return probability
expressed in terms of the diffusion propagator.

It has, however, been suggested that under certain circumstances,
ballistic trajectories (i.e., trajectories
scattering just once before returning to
their starting point) may influence the fluctuations
in the model (\ref{eq:corr},\ref{eq:H}) \cite{Mir00}.
At present it is unclear 
to which extent
the fluctuations in (\ref{eq:corr},\ref{eq:H})
may
differ from those obtained from the diffusive NLSM.
It is thus necessary to quantify
the significance of ballistic effects using alternative
methods. 

One possibility would be to derive a more general
NLSM for (\ref{eq:corr},\ref{eq:H}) incorporating ballistic effects
\cite{note}.
Very recently, for instance, a ballistic NLSM 
was used to discuss statistical properties of
energy levels and wave functions in ballistic quantum billiards
with surface scattering \cite{sasha}. 

In this paper, we  determine the statistical properties
of  wave-function amplitudes
in the model (\ref{eq:corr},\ref{eq:H})
using exact-diagonalisation calculations
for one-  and quasi-one dimensional ($1d$ and quasi-$1d$) 
tight-binding models with random on-site
potentials.  We  determine to which extent
the statistics are influenced by ballistic effects.
In addition  to spatially uncorrelated disordered
potentials [corresponding to (\ref{eq:corr})]
we also consider potentials with smoothly varying
spatial correlations (Eq.~(\ref{eq:sm_corr}) below). In the latter
case one expects ballistic contributions
to be suppressed.
Our results  support the view  that 
ballistic effects can be significant  in
quasi-one dimensional disordered systems.   

{\em Formulation of the problem}.
Within the diffusive NLSM, one finds that deviations from universal statistics
are parameterised by the time-integrated return probability
\cite{Mir00,uzy}
\begin{subequations}
\begin{equation*}
\quad P(\br,\br;\omega=0)\sim g^{-1}
\begin{cases}
1 & \text{in quasi-$1d$}\,, \qquad\hfill\refstepcounter{equation}(\theequation) \\
\log (L/\ell) & \text{in $2d$}\,,\qquad \hfill\refstepcounter{equation}(\theequation)\\
 L/\ell & \text{in $3d$}\, \qquad\hfill\refstepcounter{equation}(\theequation)
\end{cases}
\end{equation*}
\label{eq:return}
\end{subequations}
\mbox{}\\[-0.4cm]
where $P(\mathbf{r},\mathbf{r}^\prime;\omega)$  is
the diffusion propagator. 
It was pointed out in \cite{Mir00} that
the time-integrated return probability may
have additional contributions 
of the form 
\begin{subequations}
\begin{equation*}
\label{eq:ball}
\qquad\quad  P(\br,\br;\omega=0)
\sim
\begin{cases}
g^{-1} \log(\ell/\lambda_{\rm F}) & \text{in $2d$}\,,\, \qquad
\hfill\refstepcounter{equation}(\theequation)\\
\lambda_{\rm F}/\ell & \text{in $3d$}\,.
\qquad\hfill\refstepcounter{equation}(\theequation)
\end{cases}
\end{equation*}
\end{subequations}
\mbox{}\\[-0.4cm]
These arise from ballistic trajectories
contributing to the return probability
on small length scales.
In $3d$, the diffusive contribution
behaves as $\sim(\lambda_{\rm F}/\ell)^2$
which is much smaller than (4$b$).
The applicability of the diffusive NLSM is thus questionable,
an issue raised in \cite{Mir00,SmoA97}.
In $2d$ systems, the effects
are less drastic but can  nevertheless  be
appreciable.  
In quasi-$1d$ wires,
the ballistic contribution is of the form
(\ref{eq:ball}$b$) (when the sample is locally $3d$).
Then the ratio of (3$a$) and (4$b$) determines
whether ballistic effects are important or not.

The ballistic contributions (4$a,b$) are suppressed
in the case of a smoothly correlated
disordered potential, such as
\begin{equation}
\label{eq:sm_corr}
\begin{split}
\langle \upsilon(\br) \rangle &= 0\,,\\
\langle \upsilon(\br) \upsilon({\br'})\rangle
&= (2\pi\nu\tau)^{-1} f(|\br-\br'|/\ell_{\rm c})
\end{split}
\end{equation}
where $f(x)$ is a dimensionless smooth
function decaying on the scale $x\simeq 1$.
When the correlation length is
much larger than the Fermi wave length, $\ell_{\rm c} \gg \lambda_{\rm F}$,
ballistic contributions of the type (\ref{eq:ball}) are
negligible.

The following questions arise:
How important are ballistic contributions
of the type (\ref{eq:ball}$a,b$) in low-dimensional
disordered quantum systems? How exactly are they suppressed
when $\ell_{\rm c}$ is increased? Under which
circumstances is the diffusive NLSM applicable?

{\em Method.}
In order to answer these questions
we have performed exact diagonalisation studies
of a tight-binding version of (\ref{eq:H}) on
a cubic lattice with lattice spacing $a_0$
\begin{equation}
\label{eq:defH}
\widehat{H} =  \sum_{\br,\br'} t^{\phantom \dagger}_{\br\br'} c_\br^\dagger \
c^{\phantom\dagger}_{\br'}
+ \sum_\br {\upsilon}^{\phantom\dagger}_\br c_\br^\dagger 
c^{\phantom\dagger}_\br\,.
\end{equation}
Here $c_\br^\dagger$ and $c_\br$ are the 
creation and annihilation operators, the hopping amplitudes
are $|t_{\br\br'}|=1$ for nearest-neighbour sites
and zero otherwise. 
We consider $\delta$-correlated  potentials
\begin{equation}
\label{eq:c}
\langle\upsilon_\br \upsilon_{\br'}\rangle = (W^2/12) \delta_{\br\br'}
\end{equation}
as well as spatially smooth potentials
\begin{equation}
\langle\upsilon_\br \upsilon_{\br'}\rangle 
= 
 (W^2/12)
\exp[-|\br-\br'|^2/(2\ell_{\rm c})^2]\,.
\end{equation}
As usual, the parameter $W$ characterises the strength of the disorder.
As is well known, the eigenvalues $E_j$ and
wave functions  $\psi_j(\br)$ of this Hamiltonian, for
$d \geq 2$, on small energy scales and for sufficiently weak disorder
($g\rightarrow\infty$), exhibit fluctuations described by RMT. Depending on the
phases of the hopping amplitudes $t_{\br\br'}$, Dyson's Gaussian orthogonal
or unitary ensembles are appropriate \cite{meh67}. We refer
to these cases by assigning, as usual,
the parameter $\beta=1$ to the former
and $\beta = 2$ to the latter.

By diagonalising the Hamiltonian (\ref{eq:defH})
using a  modified
Lanczos algorithm \cite{CulW85a},
we have determined the distribution function
\begin{equation}
\label{eq:defP}
f_\beta(E,\br;t) 
=\Delta\Big\langle\sum_j\!
\delta(t\!-\!|\psi_j({\mathbf{r}})|^2V)g_\eta(E\!-\!E_j)\Big  \rangle
\end{equation}
of wave-function amplitudes
for $1d$ and quasi-$1d$ metallic samples.
The wave functions are normalised
so that $\langle |\psi_j(\br)|^2\rangle = V^{-1}$
and $g_\eta(z)$ is a window function of width
$\eta$, centered around $z=0$ and normalised to unity.

\begin{figure}
\psfrag{a}{(a)}
\psfrag{b}{(b)}
\psfrag{lft}{$\log_{10} f(E;t)$}
\psfrag{lt}{$\log_{10} t$}
\psfrag{lc0000}{$\ell_c/a_0=0$}
\psfrag{lc0075}{$\ell_c/a_0=0.75$}
\psfrag{lc0100}{$\ell_c/a_0=1.0$}
\psfrag{xi}{$\log_{10} (\xi/a_0)$}
\psfrag{lc}{\hspace*{4mm}$\ell_{\rm c}/a_0$}
\centerline{\includegraphics[clip,width=6cm]{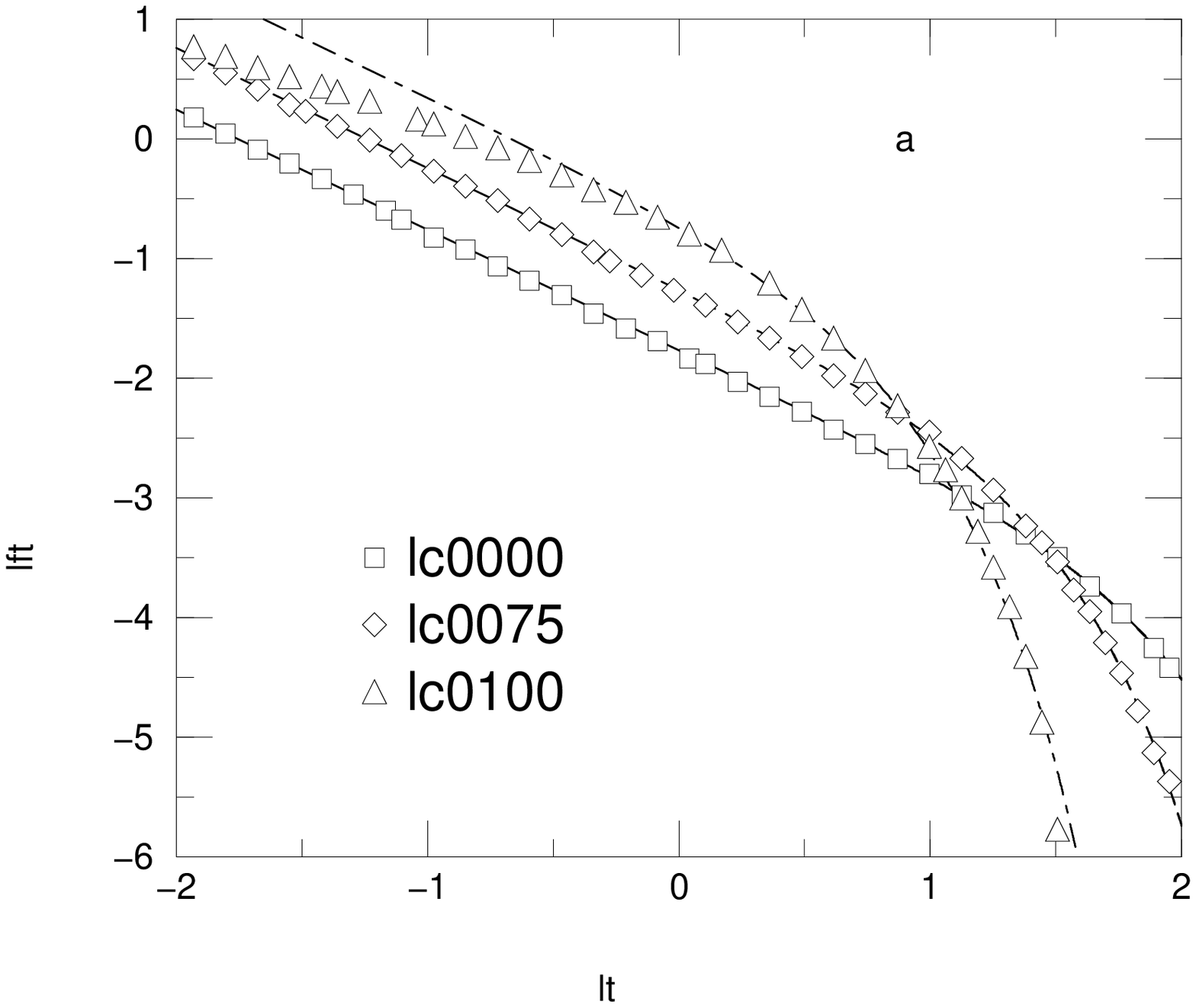}}
\centerline{\includegraphics[clip,width=6cm]{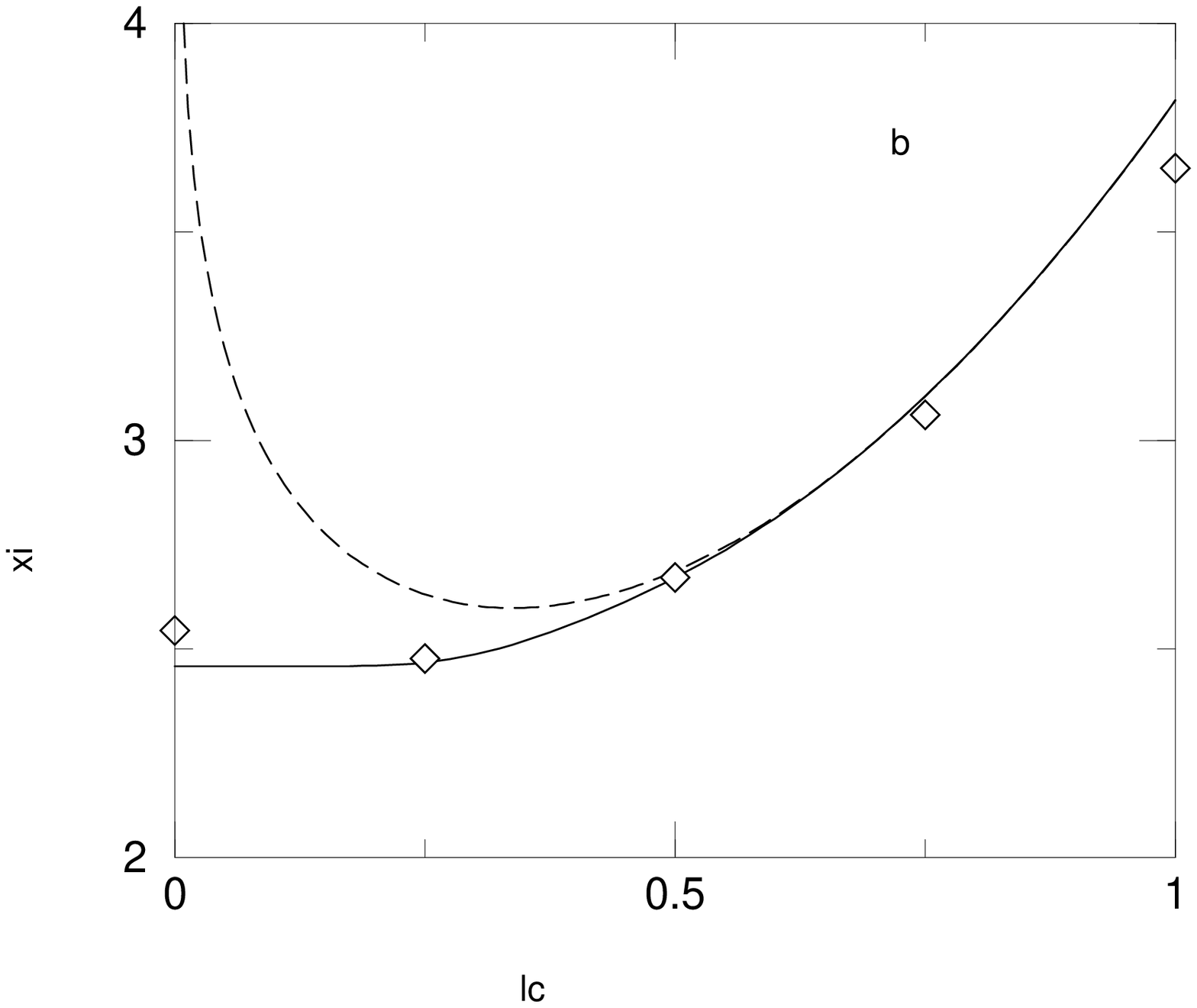}}
\caption{\label{fig:q1d} (a) The distribution
functions $f(E;t)$ for a $1d$ chain 
of length $L/a_0=2\times 10^4$, with $W=0.5$, for different values
of $\ell_{\rm c}/a_0$ (symbols). They have been fitted to Eq.\ (\ref{eq:1da}) (lines),
using $\xi/a_0$ as a fitting parameter.
(b) The corresponding localisation lengths
$\xi$ as a function of $\ell_{\rm c}$.
Also shown are expressions (\ref{eq:scba})
(full line)
and (\ref{eq:poisson}) (dashed line).}
\end{figure}

{\em Results}. We first discuss results
for a chain of length $L=2\times 10^4 a_0$
with periodic boundary conditions,
using the smoothly correlated potential (\ref{eq:sm_corr}).
Fig.~\ref{fig:q1d}(a)  shows the $x$-averaged
distribution functions
$f(E;t) \equiv \langle f(E,x;t)\rangle_x$ 
as a function of $t$ for $W=0.5,\ E=-1$ and
for several different values of $\ell_{\rm c}$.
Our numerical results are fitted
by the  expression
\begin{equation}
\label{eq:1da}
f(E;t)\simeq\frac{\xi}{L t} 
\exp\Big(-\frac{ t\, \xi}{L}\Big)
\end{equation}
derived in \cite{AltP89} for a $1d$ chain
of length $L \gg \xi$. Here $\xi$ 
is the localisation length. Eq. (\ref{eq:1da})
describes the exact-diagonalisation results
well for $t \gg L/\xi\exp(-L/\xi)$, as expected \cite{UskMRS00}.
Fig.~\ref{fig:q1d}(b) shows $\xi$ as a function
of $\ell_{\rm c}$. We observe that $\xi$ increases
exponentially fast with $\ell_{\rm c}$ for $\ell_{\rm c}$
not too large. This is in keeping
with second-order perturbation theory \cite{IzrK99},
valid for small $W$ where
\begin{equation}
\label{eq:scba}
\xi^{-1} = \frac{W^2}{96\,a_0} \frac{1}{1-E^2/4}
\sum_{x=-\infty}^\infty {\rm e}^{\ds -x^2a_0^2/(2\ell_{\rm c})^2}
{\rm e}^{\ds 2\ri x k}
\end{equation}
with $E = 2\cos(k)$.
Evaluating the sum over $x$ using Poisson summation one
obtains for $E=-1$ and large $\ell_{\rm c}$
\begin{equation}
\label{eq:poisson}
\xi^{-1} \simeq
\frac{W^2}{36\,a_0^2} 
\sqrt{\pi}
\ell_{\rm c} 
\exp[-4\pi^2\ell_{\rm c}^2/(9a_0^2)]\,.
\end{equation}
We observe reasonable agreement between
Eqs. (\ref{eq:scba}) (and (\ref{eq:poisson}) for large $\ell_{\rm c}$) and the
results of our simulations.
For stronger disorder and shorter chains
we observe significant deviations from (\ref{eq:scba}) (not shown).

We now discuss our results for quasi-$1d$ wires.
 As is well known, in the limit of large dimensionless conductance,
 the distribution function (\ref{eq:defP}) tends to  the
 RMT result 
 \begin{equation}
 f_\beta^{(0)}(t) = 
  \left\{
  \begin{array}{ll}
  \exp(-t/2)/\sqrt{2\pi t} & \mbox{for $\beta = 1$\,,}\\
  \exp(-t) & \mbox{for $\beta = 2$\,.} 
  \end{array}
  \right .
 \end{equation}
 The $\beta\!=\!1$-distribution is
 often referred to as the Porter-Thomas
 distribution \cite{por65}.
Within the diffusive NLSM one obtains for the
$x$-averaged 
relative deviations 
$\delta f_\beta(E;t) \equiv \langle f_\beta(E,x;t)\rangle_x/f_\beta^{(0)}(t)-1$
 \begin{eqnarray}
 \label{eq:q1ddev}
 \delta f_\beta(E;t)  \!&\simeq &\!
 P\left\{
 \begin{array}{ll}
 3/4\!-\!3t/2\!+\!t^2/4 &\!\mbox{for$\,\beta\!=\!1$}\,,\\
 1\!-\!2t\!+\!t^2/2   &\!\mbox{for$\,\beta\!=\!2$}
 \end{array}
 \right .
 \end{eqnarray}
(see \cite{Mir00} and references therein).
 Here $P \equiv \langle P(x,x;\omega=0)\rangle_{x}$.
 Eq. (\ref{eq:q1ddev}) is expected to be valid
 for small values of $t$  ($ t\lesssim P^{-1/2} $).
 We emphasise that, according to the diffusive NLSM,
 $P$ is the same for $\beta=1$ and $2$:
 it  is related to the diffusion propagator,
 a classical quantity. Interference effects
 are contained in the coefficients of the polynomials
 in $t$. 

 According to Refs.~\cite{Efe97,EfeL83},
 the diffusive NLSM is expected to be valid under the following conditions
 \begin{equation}
 \label{eq:cond}
 1\ll \ell/\lambda_{\rm F} \ll S/\lambda_{\rm F}^2 
 \end{equation}
 where $S=M a_0^2$ is the
 cross-section of the wire.

How may ballistic contributions, such
as those given by Eq. (\ref{eq:ball}) 
affect the prediction (\ref{eq:q1ddev})?
The heuristic argument of Ref. \cite{Mir00}
proceeds as follows: in the $\delta$-correlated case,
short ballistic trajectories contributing to the return probability
scatter only once before returning
and are thus self-retracing. When fitting
$P$ independently for $\beta=1$ and $2$ to Eq. (\ref{eq:q1ddev}),
the values $P_\beta$ may differ. In the limit
where ballistic effects dominate,
$P_1/P_2 \rightarrow 1/2$. If ballistic
effects are absent (for sufficiently smoothly correlated $\upsilon_\br$) one expects $P_1/P_2 = 1$.

Comparing expressions (\ref{eq:return}$a$) and (\ref{eq:ball}$b$), one
sees that in the
$\delta$-correlated case
the diffusive contribution is dominant only for small
enough values of $g$, i.e., when $g\ll \ell/\lambda_{\mathrm{F}}$.
Since $g=\xi/L\simeq S\ell/(\lambda_{\mathrm{F}}^2L)$, this implies a condition
\begin{align}
\label{eq:pred}
S/\lambda_{\rm F}^2 & \ll L/\lambda_{\rm F}\\
\intertext{or, if $\lambda_{\rm F}\simeq a_0$,}
M & \ll L/a_0\,.
\end{align}
Therefore, we expect the ballistic effects to dominate for
large $M$ and small $\ell_{\rm c}$.

\begin{figure}
\psfrag{lt}{$\log_{10} t$}
\psfrag{df}{$\delta f_\beta(E;t)$}
\psfrag{M=7}{{\small $\sqrt{M}=7$}}
\psfrag{M=9}{{\small $\sqrt{M}=9$}}
\psfrag{M=10}{{\small $\sqrt{M}=10$}}
\psfrag{M=11}{{\small $\sqrt{M}=11$}}
\psfrag{M=12}{{\small $\sqrt{M}=12$}}
\psfrag{M=13}{{\small $\sqrt{M}=13$}}
\psfrag{M=14}{{\small $\sqrt{M}=14$}}
\psfrag{M=16}{{\small $\sqrt{M}=16$}}
\centerline{\includegraphics[clip,width=8cm]{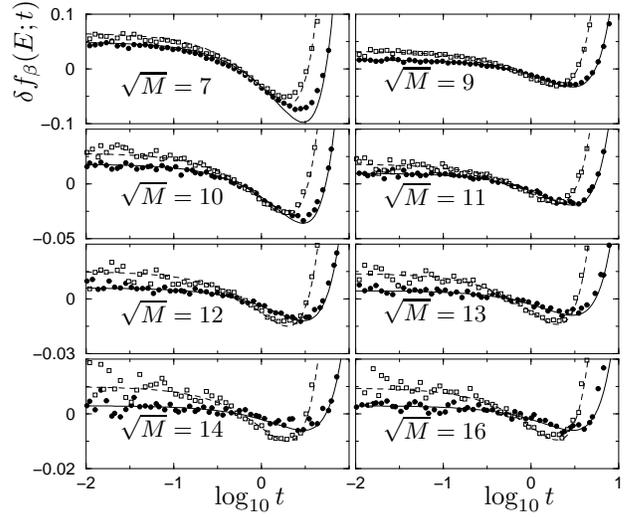}}
\caption{\label{fig:co} 
Shows  $\delta f_\beta(E;t)$ for the model (7), for $\beta=1$ ($\bullet$) and $\beta=2$
($\Box$), for $W \simeq 0.7$, $L=128\,a_0$, 
and for different values of $M$. 
Also shown are fits to the data according to (\ref{eq:q1ddev}) (lines), 
with $P$ as a fitting parameter.  
}
\end{figure}

\begin{figure}
\psfrag{lt}{$\log_{10}t$}
\psfrag{df}{$\delta f_\beta(E;t)$}
\psfrag{M=8}{{\small $\sqrt{M}=8$}}
\psfrag{M=10}{{\small $\sqrt{M}=10$}}
\psfrag{M=11}{{\small $\sqrt{M}=11$}}
\psfrag{M=12}{{\small $\sqrt{M}=12$}}
\psfrag{M=13}{{\small $\sqrt{M}=13$}}
\psfrag{M=14}{{\small $\sqrt{M}=14$}}
\centerline{\includegraphics[clip,width=8cm]{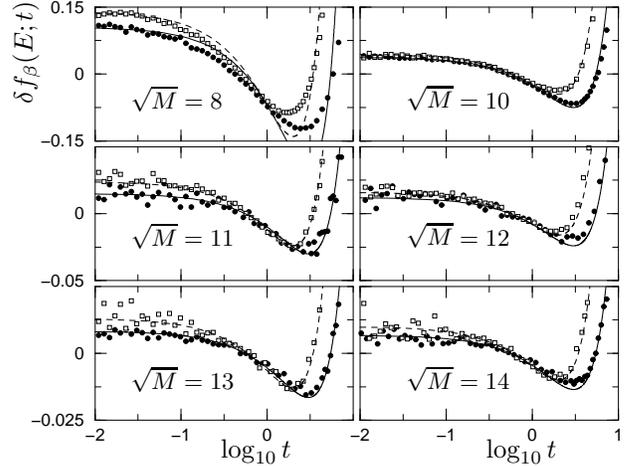}}
\caption{\label{fig:smooth}
The same as Fig.~\ref{fig:co} but for the model (8), for $\beta=1$ ($\bullet$) and
$\beta=2$ ($\square$), $\ell_{\rm c}/a_0 = 2.0$, and
$W=0.5$.  }
\end{figure}

\begin{figure}
\psfrag{P12}{$P_1/P_2$}
\psfrag{M}{$\sqrt{M}$}
\centerline{\includegraphics[clip,width=6cm]{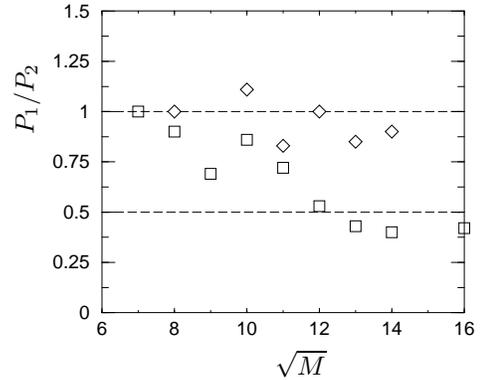}}
\caption{\label{fig4}
$P_1/P_2$ versus $\sqrt{M}$ for $\delta$-correlated ($\protect\square$) and smooth ($\protect\diamond$) potentials. 
The parameters are as in Figs. \ref{fig:co} and \ref{fig:smooth}.}
\end{figure}

 We have evaluated
 $\delta f_\beta(E,t)$ for
 $\delta$-correlated as well as for
 smooth disordered potentials in a $L\times \sqrt{S} \times \sqrt{S}$ sample
 of length $L$ and cross section $S$. 
 Our results for the $\delta$-correlated case
 are summarised in Fig.\ \ref{fig:co}. 
 The results for the smooth case ($\ell_{\rm c}/a_0 = 2$)
 are similar and are  shown in Fig.\ \ref{fig:smooth}. 
 In both cases, we have also plotted fits of Eq.~(\ref{eq:q1ddev}) 
 to the data, using the parameters $P_1$ and $P_2$ as a fitting parameters.
 We see from Figs.\ \ref{fig:co} and \ref{fig:smooth}
 that the functional forms of the deviations $\delta f_\beta(E;t)$ are
 well described by  (\ref{eq:q1ddev}).
 In Fig.\ \ref{fig4}, the values of $P_1/P_2$ are shown,
 as obtained from fitting (\ref{eq:q1ddev}) to the
 data in Figs. \ref{fig:co} and \ref{fig:smooth}, for $\beta = 1$ and $2$.
 Despite the fact that the scatter is large, we identify the following trends. 

First, in
the smoothly correlated case, $P_1/P_2\simeq 1$
for all values of $M$. This  corresponds
to the result derived using the diffusive NLSM.
In the case of a smooth potential, large-angle scattering of the
electrons is reduced and the contribution of
ballistic, self-retracing trajectories to the time-integrated
return probability is negligible.

Second, in the $\delta$-correlated case we observe
a crossover from $P_1/P_2 \simeq 1$ for small values
of $M$ to $P_1/P_2 \simeq 1/2$ for large values of $M$. The crossover
takes place at values of $M$ of the order of $L/a_0$.
This is consistent with the heuristic prediction
(\ref{eq:pred}) and indicates that
ballistic contributions to the return probability
are of the form (4$b$). They dominate the diffusive
contribution (3$a$) for large values of $M$
and are negligible for small values of $M$.
In the latter case the diffusive NLSM applies, in the
former case it does not.

{\em Conclusions}.
Using exact-diagonalisation calculations of the 
quasi-$1d$ tight-binding model (6-8), we have
determined statistical properties
of wave-function amplitudes and
compared the results to predictions of the diffusive NLSM.
Our results depend on the nature
of the spatial correlations
of the disordered potential which may
be $\delta$-correlated or spatially smooth.
The conclusions can be summarised in three points.
First, in the $\delta$-correlated case
the diffusive NLSM applies provided the wires are
not too thick ($M \ll L/a_0$). This is consistent
with the following heuristic ansatz: 
The classical return probability is
assumed to be a sum of diffusive and ballistic contributions \cite{note2}.
If $M \ll L/a_0$, the diffusive contribution
(3$a$) dominates and the diffusive NLSM is applicable.
If $M \gg L/a_0$, the ballistic contribution (4$b$)
dominates and the diffusive NLSM is not applicable
(we find $P_1/P_2 \simeq 1/2$ and not
$1$ as predicted by the diffusive NLSM).
Second, the {\em functional form of the deviations},
is still very well approximated
by the diffusive NLSM prediction even for thick wires.
Third, in the smoothly correlated
case, the exact-diagonalisation results
agree with the NLSM predictions.

Our results confirm a heuristic discussion of ballistic effects \cite{Mir00}.
Their influence on the distribution of wave-function
amplitudes is thus qualitatively understood,
but a quantitative understanding has  not yet 
been achieved. It would be of great interest
to calculate the distribution function 
$f_\beta(E,x;t)$ using the ballistic NLSM
discussed in \cite{sasha}, and to compare
to the results summarised in the present paper.

{\em Acknowledgements}. This work was supported by
the SFB 393.

\narrowtext

\end{multicols}

\begin{thebibliography}{1}

\bibitem{LifGP88}
I.~M.~Lifshitz and L.~A. Pastur, {\em Introduction to the theory of
 disordered systems} (Wiley, New York, 1988).
\bibitem{Efe97}
K. B. Efetov, Adv. Phys. {\bf 32}, 53 (1983);
K.~B. Efetov, {\em Supersymmetry in disorder and chaos} (Cambridge Univ. Press,
  Cambridge, 1997).

\bibitem{Mir00}
A.~D. Mirlin, Phys. Rep. {\bf 326},  259  (2000).

\bibitem{meh67} M.~L. Mehta, {\em Random Matrices and the Statistical Theory of Energy
                Levels} (Academic Press, New York, 1991).
\bibitem{por65} C. E. Porter, in: {\em Statistical Theories of Spectra},
                ed: C. E. Porter (Academic Press, New York, 1965).

\bibitem{note} B. Muzykantskii and D. E. Khmelnitskii, 
Pis'ma Zh. Eksp. Teor. Fiz.  {\bf 62}, 68 (1995) [JETP Lett. {\bf 62}, 76 (1995)].

\bibitem{sasha} Ya. M. Blanter, A. D. Mirlin, and B. A. 
Muzykantskii, cond-mat/0011498 v2. The model used
here is similar but not identical to that used in \cite{note}.

\bibitem{uzy}
N. Argaman, Y. Imry and U. Smilansky,
Phys. Rev. B {\bf 47}, 4440 (1993).

\bibitem{SmoA97}
I.~E. Smolyarenko and B.~L. Altshuler, Phys. Rev. B {\bf 55},  10451  (1997).


\bibitem{CulW85a}
J. Cullum and R.~A. Willoughby, {\em {Lanczos} Algorithms for Large Symmetric
  Eigenvalue Computations, Volume 1: Theory} ({Birkh\"{a}user}, Boston, 1985).

\bibitem{AltP89}
B.~L. Altshuler and V. N. Prigodin, Zh. Eksp. Teor. Fiz. {\bf 95},  348
  (1989) [Sov. Phys. JETP {\bf 68}, 198 (1989)].

\bibitem{UskMRS00}
V.~Uski, B.~Mehlig, R.~A.~R\"{o}mer and M.~Schreiber, Phys. Rev. B {\bf 62}, 
R7699 (2000).

\bibitem{IzrK99}
F.~M. Izrailev and A.~A. Krokhin, Phys. Rev. Lett. {\bf 82}, 4062
(1999).

\bibitem{EfeL83} K.~B. Efetov and A. I. Larkin, 
Zh. Eksp. Teor. Fiz. {\bf 85}, 764 (1983) [Sov.
Phys. JETP  {\bf 58}, 444 (1983)].

\bibitem{note2} Very recently it was shown
that in ballistic quantum billiards with surface scattering,
the return probability is of this form \cite{sasha}.

\end{thebibliography}
\end{document}